\documentclass[aps,superscriptaddress,amsmath,amssymb,longbibliography,preprint,11pt]{revtex4-1}
\usepackage{amsmath,amsfonts,amssymb}
\usepackage{graphicx,subfigure}
\usepackage{epsfig}
\usepackage{soul}

\begin{document}

\title{
Convergent trajectories of relativistic electrons interacting with lasers in plasma waves 
}

\author{Bin Liu}
\affiliation{Guangdong Institute of Laser Plasma Accelerator Technology, Guangzhou, China}

\author{Bifeng Lei}
\affiliation{Department of Physics, The University of Liverpool, Liverpool, L69 3BX, United Kingdom}
\affiliation{State Key Laboratory of Nuclear Physics and Technology, School of Physics, CAPT, Peking University, Beijing, China}

\author{Matt Zepf}
\affiliation{Helmholtz Institute Jena, Fr\"{o}belstieg 3, 07743 Jena, Germany}
\affiliation{Institute of Optics and Quantum Electronics, Friedrich-Schiller-Universit\"{a}t Jena, 07743 Jena, Germany}

\author{Xueqing Yan}
\affiliation{Guangdong Institute of Laser Plasma Accelerator Technology, Guangzhou, China}
\affiliation{State Key Laboratory of Nuclear Physics and Technology, School of Physics, CAPT, Peking University, Beijing, China}
\affiliation{Beijing Laser Acceleration Innovation Center, Huairou, Beijing, China}

\begin{abstract}
The dynamics of relativistic electrons interacting with a laser pulse in a plasma wave has been investigated 
theoretically and numerically based on the classical Landau-Lifshitz  equation. 
There exists a convergent trajectory  of electrons when 
the energy gain of electrons via direct laser acceleration 
can compensate the energy loss via radiation. 
An electron beam initially around the convergent trajectory  evolves  
into the trajectory, 
making its occupied phase space volume decrease exponentially 
while mean energy remain the same. 
This mechanism can be used for cooling relativistic electron beams especially those 
produced in plasma-based acceleration.   
\end{abstract}

\maketitle

Plasma-based electron acceleration
has made significant progress in the last decades. 
Plasma waves with both longitudinal accelerating fields and  
transverse focusing fields for electrons 
can be produced as wakes of 
either laser pulses or charged particle beams \cite{Tajima1979,Chen1985,awake1,Adli2018}. 
Tens of GeV energy enhancement of electrons has been achieved via electron beam driven acceleration~\cite{Blumenfeld:2007aa}.
Electron beams with energy up to the order of 10 GeV 
have been produced via laser driven approaches \cite{Gonsalves2019,Hegelich2023}.
Besides of achieving high energy electron acceleration,
many efforts are devoted to improving the beam quality, 
which is important for 
applications ranging from   
X-ray free electron lasers \cite{Schroeder2006,Gruner2007}, 
and QED physics \cite{Poder2018,Cole2018},
to electron-positron and photon colliders \cite{Schroeder2010}.  
Electron beams with energy hundreds of MeV and 
relative energy spread down to per-mille level,
which is compatible to that in the state-of-the-art conventional accelerators, 
have been produced in several experiments \cite{Ke2021,Wang2021,Pompili2022,Galletti2022}.
Nevertheless, achieving such beam quality for electron beams with energy well beyond GeV in plasma-based acceleration remains a challenge. 

Cooling via radiation damping is an effective method for improving the quality of electron beams, as has been adopted in conventional damping rings, see, e.g., \cite{Emma2001}.
In a typical damping ring, electrons radiate energy in the field of the bending magnets and meanwhile regain energy via radio-frequency acceleration, allowing an electron beam to maintain its mean energy while reducing its emittance.
This works due to the damping feature of radiation reaction. 
Radiation damping occurs naturally in plasma-based electron acceleration 
since electrons inevitably experience betatron oscillation under the transverse focusing fields \cite{Esarey2002,Corde2013}. 
The effects of radiation damping on the electron dynamics, acceleration,  
and especially the electron beam quality 
have attracted much attention \cite{Barletta1999,Michel2006,Kostyukov2012,Shpakov2018,FerranPousa2019,Zeng2021,Golovanov2022,Deng2012}.

The transverse oscillation of 
electrons in plasma waves 
can be modulated and enhanced 
by applying an interacting laser, 
known as direct laser acceleration (DLA) \cite{Pukhov1999,Cipiccia2011}. 
This sets a configuration where electrons gain energy and radiate energy simultaneously, similar to that in a damping ring. 
In this work, we show that,  
when the interacting laser is strong enough so that the energy gain of electrons via DLA 
can compensate the energy loss via radiation, 
there exists a convergent trajectory around which an electron beam evolves into the trajectory  
with its occupied phase space volume decreasing exponentially 
and mean energy remaining the same. 
This can be exploited for 
cooling relativistic electron beams, especially those generated via 
plasma-based acceleration.

A schematic plot of the configuration is shown in Fig. 1 (a). 
An electron beam is confined in the cavity of a fast-moving plasma wave 
and interacts with an ultra-short laser pulse which comoves with the plasma wave. 
For simplicity, 
we assume that the phase velocity of the plasma wave $v_p$ is a constant,  
and the group velocity of the interacting laser $v_g$ 
equals to $v_p$, which is achievable by adjusting the interacting laser, 
the plasma, and the driving source of the plasma wave. 
We first consider a case that the interacting laser pulse is 
right-hand circularly polarized (RHCP).
We use a cylindrical coordinate system $(r,\theta,z)$,  
where $z$ is the plasma wave propagating direction. 
We focus on the interaction near the laser axis  
where we can use a plane wave approximation for 
the laser electric and magnetic fields, 
${\boldsymbol{E}}_L 
= \left(E_{Lr}, E_{L\theta}, 0 \right)
= \left(E_{L} \cos\psi , E_{L} \sin \psi, 0\right)$,  
and 
${\boldsymbol{B}}_L 
= \left(B_{Lr}, B_{L\theta}, 0 \right)
= \left(- v_g\sin \psi E_{L}, v_g\cos\psi E_{L} , 0\right)/c^2$, 
where $v_g$ is the laser group velocity, 
$\psi= k_L z - \omega_L t - \theta$  
the phase difference between the laser field and the 
transverse oscillation of the electron (Fig. 1(b)),
$k_L$ and $\omega_L$ the laser wavenumber and frequency, 
respectively.  
The self-generated
electric and magnetic fields in the  plasma wave  can be 
expressed as 
$\boldsymbol{ {E}}_{S}=( {E}_{Sr},0, {E}_{Sz})$ 
and $\boldsymbol{ {B}}_{S}=(0,  {B}_{S\theta} ,0)$, 
which can be written as functions of a rescaled coordinate $\zeta = z-v_p t$.

Since the laser pulse copropagates with the electron beam, the interaction between them is usually weak so that one can write the equation of motion of electrons in 
the classical Landau-Lifshitz approach, 
\begin{eqnarray}
\frac{d}{d t} \left(\gamma m \boldsymbol{v} \right) =
\boldsymbol{F}_{\rm{ext}}  + \boldsymbol{F}_{\rm{rad}},  
\end{eqnarray}
where, in relativistic limit $v\to c$, 
by taking the leading term, the radiation reaction force becomes  
\begin{eqnarray}
\boldsymbol{F}_{\rm{rad}}  =  
- {2 r_e F_{\perp}^2 \gamma^2 \boldsymbol{v} } / {3 m c^3}, 
\end{eqnarray}
here 
$r_e = {e^2}/{4\pi \epsilon_0 m c^2}$    
is the classical electron radius,  
$F_{\perp} = \sqrt{ \boldsymbol{F}_{\rm{ext}} ^2 - F_v^2 }$ 
and $F_v = \boldsymbol{v} \cdot  \boldsymbol{F}_{\rm{ext}} / c$ are
the Lorentz forces perpendicular to and along with
the direction of the electron velocity $\boldsymbol{v}$, respectively, 
$\boldsymbol{F}_{\rm{ext}} 
= -e (\boldsymbol{E}_{\rm{ext}} +\boldsymbol{v}\times \boldsymbol{B}_{\rm{ext}} )$ 
is the external Lorentz force with 
the electric field $\boldsymbol{E}_{\rm{ext}} =\boldsymbol{E}_L+\boldsymbol{E}_S$ 
and the magnetic field $\boldsymbol{B}_{\rm{ext}}=\boldsymbol{B}_L+\boldsymbol{B}_S$, 
$\gamma=1/\sqrt{1-v^2/c^2}$ denotes the electron Lorentz factor,  
$\epsilon_0$ the  vacuum permittivity,
$c$ the light speed in vacuum, 
$m$ and $e$ the electron mass and charge, respectively.  

By introducing the dimensionless variables,  
$\beta = v/c$,
$\tau = \omega_L t$, 
$\nu = d\theta /d \tau$,
$\mathcal{F} = F/m c \omega_L$, 
$\mathcal{E} = e E/m c \omega_L$,  
$\mathcal{B} = e B / m \omega_L$, 
$\rho = r/(c/\omega_L)$, 
$\xi = \zeta /(c/\omega_L)$, 
and 
$\delta = 2 \omega_L r_e / 3 c$, 
Eq. (1) can be rewritten as 
\begin{align}
 \frac{d}{d \tau} \gamma &=  - \beta_z \mathcal{E}_{Sz} 
- \beta_r (\mathcal{E}_{Sr}+\mathcal{E}_{Lr}) - \rho \nu \mathcal{E}_{L\theta}
 -\delta  \gamma^2 \mathcal{F}_{\perp}^2, \label{eqdg}   \\
 \frac{d}{d \tau}  \nu &= - \frac{\nu}{\gamma} \frac{d \gamma}{d \tau}
- \frac{2}{\rho} \beta_r  \nu
-\frac{\eta \mathcal{E}_{L\theta}}{\gamma \rho}  
- \delta \gamma   \nu \mathcal{F}_{\perp}^2,  
\label{eqdnu} \\
 \frac{d}{d \tau} \beta_r &= -\frac{\beta_r}{\gamma} \frac{d \gamma}{d \tau}
+ \rho \nu^2  
-\frac{\eta \mathcal{E}_{Lr}}{\gamma}  
+ \frac{\mathcal{F}_{Sr} }{\gamma} 
- \delta  \gamma \beta_r \mathcal{F}_{\perp}^2,  \label{eqdbr}   \\
 \frac{d}{d \tau} \psi &= - \nu - \eta,  \label{eqdpsi}  \\
 \frac{d}{d \tau} \rho &= \beta_r,  \label{eqdrho}  \\
 \frac{d}{d \tau} \xi  &= \beta_z - \beta_p, \label{eqdxi}
\end{align}
where  $\eta = 1 - v_g v_z / c^2 $, and  
$\mathcal{F}_{Sr} = -\mathcal{E}_{Sr}+v_z \mathcal{B}_{S\theta}$ 
is the force of the plasma wave fields in the radial direction. 
We look for 
stable fixed points of the equations at where  
the derivations of the variables, i.e., 
the left hand sides of the equations, 
are all equal to zero. 
We mark the value of a variable 
at the stable fixed point by putting a bar over it. 
Letting the right hand sides of Eqs. (\ref{eqdpsi},\ref{eqdrho},\ref{eqdxi}) be zero, we have 
\begin{eqnarray}
\bar{\nu} &=   - \bar{\eta},    
\label{eqnu} \\
\bar{\beta}_{r} &=0,  
\label{eqbr} \\
\bar{\beta}_z &= \beta_p.  
\label{eqbz}
\end{eqnarray}
By making use of Eq. (11) and $\beta_g=\beta_p$, one gets 
\begin{eqnarray}
\bar{\eta}  =  
 1/\gamma_p^2, 
\label{eqeta}
\end{eqnarray}
where $\gamma_p = 1/\sqrt{1-\beta_p^2}$. 
Furthermore, according to 
$\bar{\gamma} = 1/\sqrt{1-(\bar{\beta}_{z}^2 + \bar{\beta}_{r}^2 + \bar{\rho}^2\bar{\nu}^2)}$ and 
Eqs. (\ref{eqnu}-\ref{eqeta}), 
in the limit of $\bar{\gamma} \gg \gamma_p$, one has   
\begin{eqnarray}
\bar{\rho} & =  \gamma_p.  
\label{eqrho} 
\end{eqnarray}
By making use of Eq. (\ref{eqbr}) 
and letting the right hand sides of Eqs. (\ref{eqdg}) and (\ref{eqdnu}) 
be zero,   one gets 
\begin{align}
&  - \bar{\beta}_z \bar{\mathcal{E}}_{Sz} -\bar{\rho} \bar{\nu} \bar{\mathcal{E}}_{L\theta} - \delta \bar{\gamma}^2 
\bar{\mathcal{F}}_{\perp}^2 = 0, \label{eqb1}  \\
& -\bar{\eta} \bar{\mathcal{E}}_{L\theta}
- \delta \bar{\gamma}^2 \bar{\rho}  \bar{\nu}  \bar{\mathcal{F}}_{\perp}^2 =0,  \label{eqb2}  
\end{align} 
By combining them and making use of Eqs. (\ref{eqnu},\ref{eqeta},\ref{eqrho}),  
one obtains  
\begin{equation}
\bar{\mathcal{E}}_{Sz} =  0.  
\end{equation}
This indicates that the electron beam tends to be accumulated to the  
center of the plasma wave cavity in the longitudinal direction  
where the longitudinal wakefield vanishes
and therefore the assumption $\beta_g = \beta_p$ can be satisfied and maintained locally 
even after long propagation \cite{Wilks1989,Esarey2009}.
Then one has 
$\bar{\mathcal{F}}_v = -v_{\theta} \mathcal{E}_{L\theta} 
= - \mathcal{E}_{L} \sin\bar{\psi}/\gamma_p $ 
and thus 
$\bar{\mathcal{F}}_{\perp} 
= \bar{\mathcal{F}}_{Sr} -  \mathcal{E}_{L}  \cos \bar{\psi} /\gamma_p^2$. 
Furthermore, according to Eqs. (\ref{eqdbr}) and (\ref{eqbr}), one gets  
\begin{equation}
\bar{\rho} \bar{\nu}^2 -\frac{\bar{\eta} \bar{\mathcal{E}}_{Lr}}{\bar{\gamma}} + \frac{\bar{\mathcal{F}}_{Sr}}{\bar{\gamma}} =0,  
\label{eqb3} 
\end{equation}
By combining Eqs. (\ref{eqb2}) and (\ref{eqb3}), 
and introducing $\mathcal{D} = \delta \gamma_p^5 |\bar{\mathcal{F}}_{Sr}|^3$ 
and $\mathcal{A} = \mathcal{E}_L/\left(\gamma_p^2  |\bar{\mathcal{F}}_{Sr}| \right)$, 
we have the equation of $\bar{\psi}$, 
\begin{equation}
\mathcal{A}  \sin\bar{\psi} - \mathcal{D} 
(\mathcal{A} \cos \bar{\psi} + 1)^4  =0. 
\end{equation}
We focus on the condition of $\mathcal{A}\ll 1$, 
under which there are two solutions, 
${\psi}_1 = \tan^{-1} ( (1+4 \mathcal{M})\mathcal{D}/(\mathcal{M}-4\mathcal{D}^2) )$, 
$ {\psi}_2 =  \pi - \tan^{-1} ( (1-4 \mathcal{M})\mathcal{D} /(\mathcal{M}+4\mathcal{D}^2) )$, 
where 
$\mathcal{M}=\sqrt{\mathcal{A}^2(1+16\mathcal{D}^2)-\mathcal{D}^2}$ 
gives a restriction condition for having real solutions,  
\begin{equation}
\mathcal{A}>\mathcal{D}/\left(\sqrt{1+16\mathcal{D}^2 }\right). 
\end{equation}
This corresponds to the condition that the laser field  
is strong enough so that the energy gain of electrons from the laser 
can compensate the energy loss via radiation. 
It is challenging to investigate the stability of the solutions analytically.
Numerically, we have found that ${\psi}_1$ 
is stable and ${\psi}_2$ is unstable. 
The former solution corresponds to a 
physical configuration where the laser force along the radial direction pulls the electron inward,
while for the latter one the laser force pushes the electron outward. 
This coincides with the results obtained from the simplified models in Refs. \cite{Liu2015,Hu2015,Gong2018}. 
Therefore, we have 
\begin{align}
\bar{\psi} &=  \tan^{-1} \left( \frac{(1+4 \mathcal{M})\mathcal{D} }{\mathcal{M}-4\mathcal{D}^2} \right),   \label{barpsi}  \\
\bar{\gamma} &= 
\frac{1+\mathcal{M} +12 \mathcal{D}^2}{\mathcal{A} (1+16\mathcal{D}^2 )} \gamma_p \mathcal{E}_L.   \label{bargamma} 
\end{align}

It is shown that the convergent trajectory corresponds to a helical curve in real space with 
a fixed radius $\gamma_p c /\omega_L$ and 
a constant phase difference $\bar{\psi}$ with respect to the laser field. 
In this RHCP case, 
the balance between the energy gain and the radiation loss is achieved  
all the time for electrons at the convergent trajectory. 
Actually,   
Eqs. (\ref{eqb1}), (\ref{eqb2}) and (\ref{eqb3}) 
correspond to the balance of the instantaneous forces    
in the  direction of $\boldsymbol{v}$,  
$F_v+F_{\rm{rad}} = 0$, 
in $\theta$-direction, 
$F_{L\theta}+F_{\rm{rad}} \sin \phi =0$, 
and in $r$-direction,  
$F_{Lr}+F_{Sr}+\gamma m v_{\theta}^2/r = 0$,  
respectively, 
as is shown in Fig. 1 (b) and (c), 
where $\phi$ denotes the included angle 
between $\boldsymbol{v}$ and $z$-direction. 
Furthermore, in the frame comoving with the plasma wave, 
the convergent trajectory becomes a closed circle. 
Therefore, 
the configuration presented here can be seen as 
a moving micro damping ring with radius $\gamma_p c /\omega_L$ and moving speed $v_p$.

In order to see more details of the cooling process, 
the Landau-Lifshitz equation has been solved numerically. 
The interacting laser is assumed to be a RHCP  
plane wave laser with laser amplitude $\mathcal{E}_L=0.5$ and wavelength $\lambda_L= 400 \rm{n m}$, corresponds to $\delta = 2.95\times 10^{-8}$. 
One can increase the cooling efficiency by increasing the electric field of the interacting laser 
although more complexity may arise. 
For the plasma wave, we assume $\gamma_p=100$ so that 
the phase velocity of the plasma wave is $\beta_p = 0.99995$. 
Without loss of generality, 
the plasma wave fields near the longitudinal center of the plasma wave cavity  
($\zeta = 0 $ where $E_{Sz} = 0$) are approximated as 
$E_{Sz} = k_1  m \omega_L^2  \zeta / e$, 
$ E_{Sr} = k_2  m \omega_L^2  r / e$, 
and 
$ B_{S\theta} = - k_3 m \omega_L^2 r /e c $,
with    
$k_1 = k_2 = k_3 = 7.5\times 10^{-5}$. 
Then, one obtains $\bar{\rho}=100$, $\bar{\nu}=10^{-4}$, 
$\bar{\xi}=\bar{\beta}_r=0$,
$\bar{\psi} = 0.30727$ and $\bar{\gamma} = 15047.283$. 
It is noticed that   
the quantum radiation effect \cite{Piazza2010,Piazza2012,Neitz2013,Vranic2016,Harvey2017} 
is neglectable since 
$\bar{\gamma} \bar{F}_{\perp}/e E_c = \bar{\gamma}^2 m c \omega_L 
/ (\gamma_g^3 e E_c) \approx 10^{-3} $, 
where 
$E_c= m^2 c^3/e \hbar$  
is the Schwinger limit of electric field. 
Totally $N=10^5$ randomly selected test electrons have been calculated 
in a Cartesian coordinate system ($x,y,z$) 
with the variables initially normally distributed  
around a point at the convergent trajectory 
($\langle x \rangle =\bar{\rho} c/\omega_L$, 
$\langle y \rangle = \langle z \rangle=
\langle p_{x} \rangle = 0$, $\langle p_{y} \rangle =-\bar{\gamma}/\gamma_p $, $\langle p_{z} \rangle=\bar{\gamma} \beta_p$)  
with standard deviations  
$\sigma(x) = \sqrt{\langle x^2 \rangle} = c/\omega_L $,
$\sigma(y) = c/\omega_L $, 
$\sigma(z) = 0.3 c/\omega_L $, 
$\sigma(p_x)  = \sigma(p_y)   = 1.5$
and $\sigma(p_z)= 150$, 
where $x=r\cos\theta$, $y=r\sin\theta$, $p_x=\gamma \beta_x$, 
$p_y=\gamma \beta_y$, and $p_z=\gamma \beta_z$ 
\footnote{
For an electron beam with much larger phase space volume, 
the electrons initially near the stable fixed point 
with $\psi \sim \psi_1$ 
can be cooled to the convergent trajectory, 
similar to that in Fig. 2,  
while others initially near the unstable fixed point with 
$\psi \sim \psi_2$ may spread out 
and lose energy quickly, 
as is demonstrated in the Supplemental Material at [URL will be inserted by publisher].   
}.
Fig. 2 (a-c) shows the initial distributions of the electrons 
in the phase spaces 
$\rho-p_r$, $\theta-p_\theta$, and $\xi-p_z$, 
re-centered to the corresponding values at the convergent trajectory 
except for $\theta$, 
respectively, 
where $p_r=\gamma \beta_r$ and $p_{\theta} = \gamma \rho^2 \nu $. 
The corresponding distributions after 20 ns are shown in 
Fig. 2 (e-g) in which the electrons are much more close to the convergent trajectory, 
resulting in a significant reduction of the occupied 
phase space area of the electron beam.  
The distributions of the electrons in the space spanned by 
$\psi-\gamma$, re-centered to ($\bar{\psi},\bar{\gamma}$), 
at $t=0$ and $t=20$ ns 
are shown in Fig. 2 (d) and (h), respectively, 
confirming that the theoretical result of the convergent trajectory works well.

In order to have a comprehensive understanding,   
the evolution of the occupied phase space area 
in different directions is shown in Fig. 3. 
Since the distribution of the electrons in each phase space 
is always elliptical-like, 
the occupied phase space area is calculated as 
the determinant of the covariance matrix of the 
phase space coordinates, 
$\mathcal{S}_i = \sqrt{\left\langle q_i^2 \right\rangle \left\langle p_{i}^2 \right\rangle - \left\langle  q_i  \cdot p_{i} \right\rangle^2 } $, 
where 
$i =r,\theta , z$ 
corresponds to  the phase space  
$\rho - p_{r}$, $\theta- p_{\theta}$, and $\xi -p_z$, 
respectively. 
It is seen that $\mathcal{S}_r$ drops almost monotonically (Fig. 3 (a)), 
indicating that the electrons are well confined in the radial direction.
The long term evolution can be well fitted by an exponential decay 
$\mathcal{S}_r/\mathcal{S}_{r0} = \exp(-t/T_r)$, where 
the lifetime is $T_r = 4.2$ ns.  
On the other hand, electrons spread out quickly 
in $\theta$ direction at the beginning (in 0.1 ns) 
\footnote{See Supplemental Material at [URL will be inserted by publisher] for more details.},  
resulting in a sudden increase in $\mathcal{S}_{\theta}$ 
by a few hundredfold. 
This suggests that the phase space structure near 
the convergent trajectory may be complex. 
A more dedicated research is required. 
Nevertheless, afterwards, 
the electron beam in the phase space 
shrinks in both $\theta$ and $p_{\theta}$ directions. 
Although strong fluctuations exist at the early stage ($t<8$ ns), 
the overall evolution of $\mathcal{S}_{\theta}$ 
can still be approximated as an exponential decay, 
$\mathcal{S}_{\theta}/\mathcal{S}_{\theta 0} = f_{\theta} 
\exp(-t/T_{\theta})$, 
as is seen in Fig. 3 (b), 
where $f_{\theta}=100$ characterizes the spread rate at the beginning, 
and the lifetime is fitted as $T_{\theta} = 2.2$ ns, 
indicating a higher decay rate than that 
in the radial direction. 
In $z$ direction, the situation is similar to that in 
$\theta$ direction and one has the fitting  
$\mathcal{S}_z / \mathcal{S}_{z 0}  =  f_z\exp(-t/T_z)$ with 
the same lifetime $T_z =  2.2$  ns, 
except that the spread rate at the beginning $f_z=10$ 
is one order of magnitude lower (Fig. 3 (c)).

Although the discussion above is for RHCP lasers, 
it is clear that for left-hand circularly polarized lasers the result is exactly the same except that the direction of rotation of the helical trajectory is opposite. 
For linearly polarized lasers, 
convergent trajectories still exist
as long as the  energy gain via DLA  
and the energy loss via radiation 
cancel each other out in betatron cycles. 
For instance, by using a LP laser with the same laser intensity 
and the same setup used in Fig. 2,  
we have observed similar efficient cooling process 
of the electron beam, 
although the mean energy keeps oscillating 
around $\gamma \sim 15027$ even after 30 ns.   

In summary, there exists a convergent trajectory  
for relativistic electrons interacting with a laser pulse  in a plasma wave  
as long as the laser field is strong enough so that 
the energy gain of the electrons via DLA 
can compensate the energy loss via radiation. 
For an electron beam initially near the convergent trajectory, 
its occupied phase space volume reduces exponentially over time 
while the mean energy is maintained, resulting in high efficient cooling of the electron beam. 
This plasma-based cooling configuration is featured by high efficiency and relatively small size, 
making it especially suitable for improving the quality of electron beams from plasma-based accelerators, which is important for various applications 
including X-ray free electron lasers  \cite{Schroeder2006,Gruner2007}, 
QED physics \cite{Poder2018,Cole2018}, and colliders \cite{Schroeder2010}.

Bin Liu acknowledges the support of Guangdong High Level Innovation Research Institute Project, 
Grant No. 2021B0909050006. 
Bifeng Lei acknowledges the support of the Science and Technology on Plasma Physics Laboratory, Grant No. 6142A04210110. 
Xueqing Yan was supported by the NSFC (Grants No. 11921006,), Beijing Outstanding Young Scientists Program, the National Grand Instrument Project (No.2019YFF01014400).


\begin{figure}[h]
\centering
\includegraphics[width=0.45\textwidth]{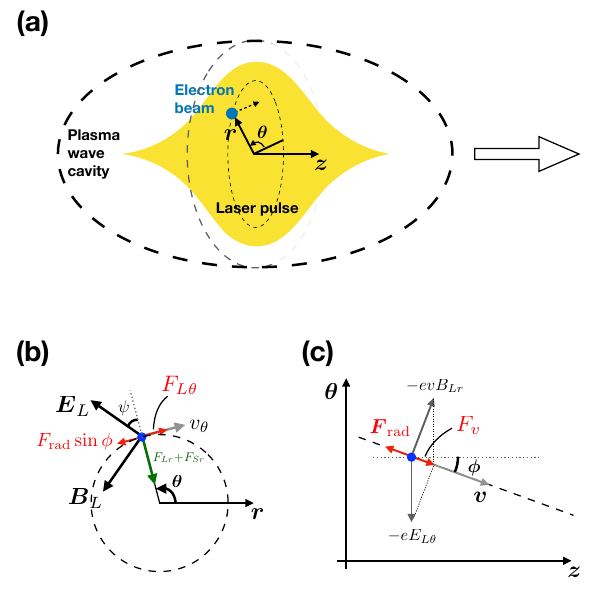}
\caption{
Schematic plot. 
(a) A relativistic electron beam (blue disk) 
interacting with a circularly polarized 
laser pulse (yellow region) 
in a plasma wave cavity which can be driven by either a laser pulse or a charged particle beam (not plotted). 
When the energy gain of electrons via direct laser acceleration
can compensate the energy loss via radiation, 
there exists a convergent trajectory around where  
the cooling of the electron beam occurs. 
Instantaneous forces on electrons at the convergent trajectory,
which is a helical curve with both $\psi$ and $\phi$ fixed, 
in (b) $(r-\theta)$ and (c) $(z-\theta)$ projections, 
where  $\psi$ denotes the phase difference between the laser electric field
and the transverse oscillation, 
and $\phi$ the included angle between 
the movement direction $\bm{v}$ and $z$-direction. 
}
\label{fig1}
\end{figure}

\begin{figure}[t]
\centering
\includegraphics[width=0.45\textwidth]{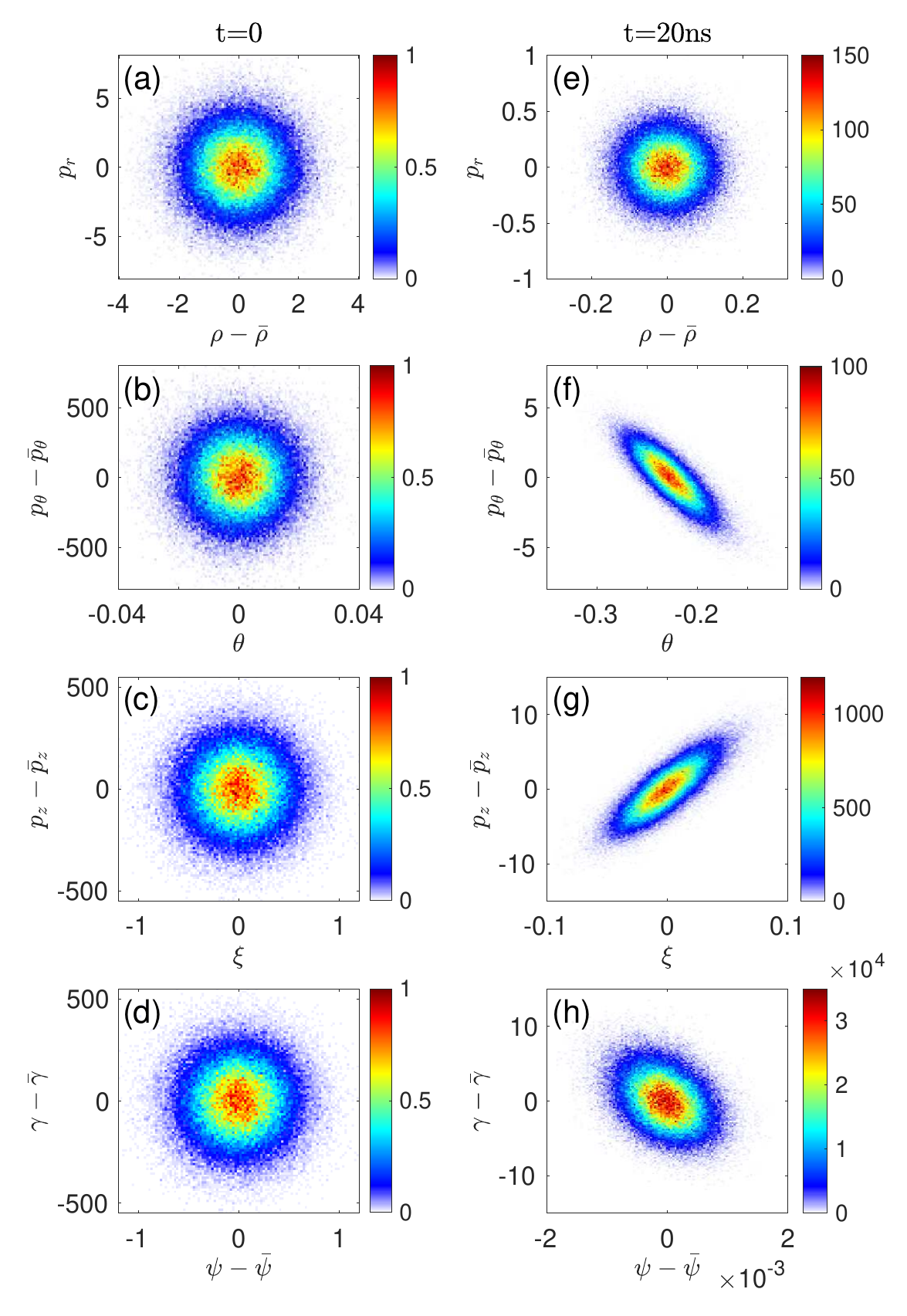}
\caption{
Distributions of electrons in phase space 
(a,e)   
$\rho-p_r$,  
(b,f) $\theta-p_{\theta}$, 
(c,g) $\xi-p_z$, 
and (d,h) the space spanned by $\psi-\gamma$,  
re-centered to the values at the convergent trajectory  
(with bars over them) 
except for $\theta$, at 
(a-d) $t=0$ and 
(e-h) $t= 30$ ns, 
obtained by numerically solving the Landau-Lifshitz equation 
for $10^5$ electrons initially near the convergent trajectory 
(see text for more detail). 
Notice the different scales for the left and right panels. 
}
\label{fig3}
\end{figure}

\begin{figure}[b]
\centering
\includegraphics[width=0.45\textwidth]{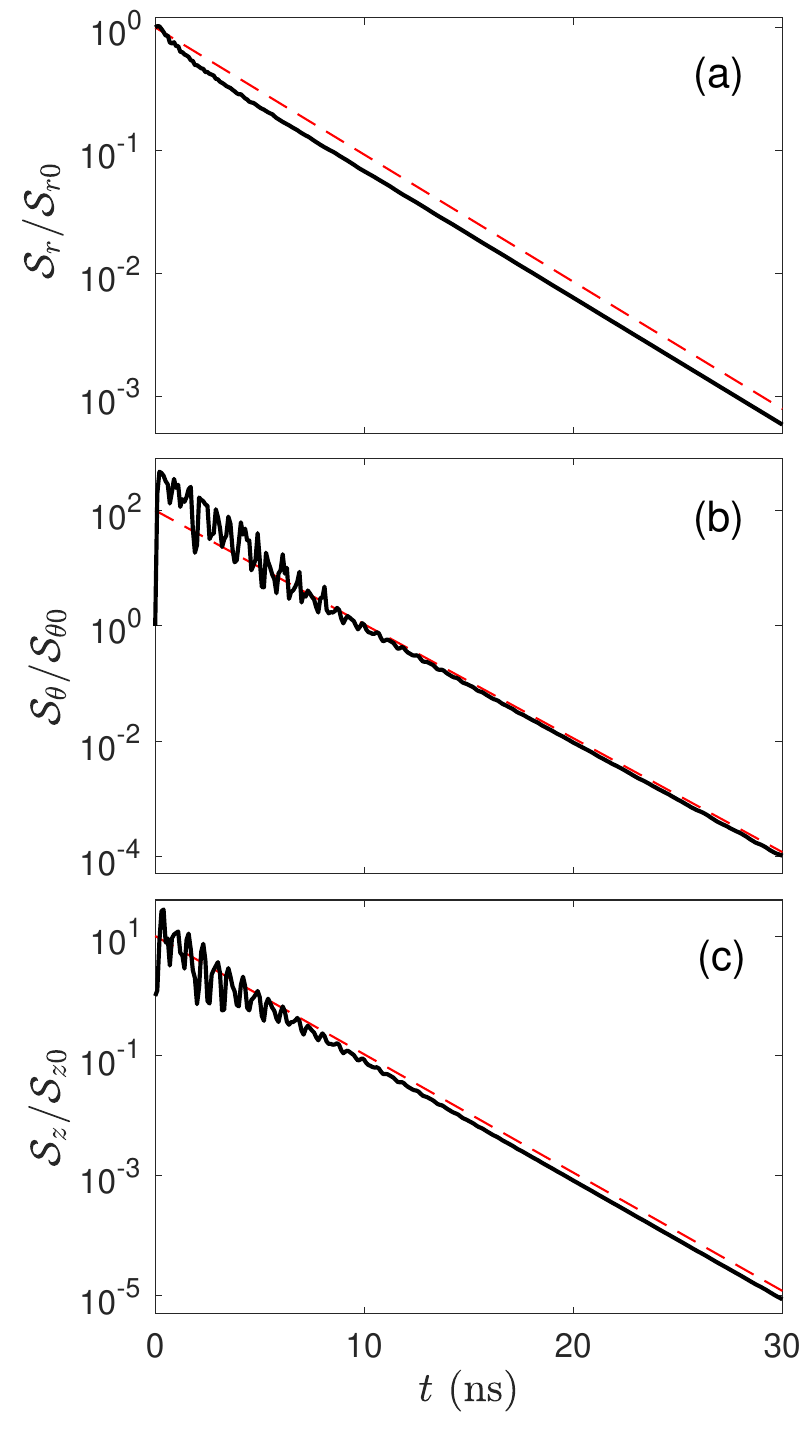}
\caption{
Evolutions of 
the occupied area in the phase space in different directions (black solid curves),  
(a) $\mathcal{S}_r$ for phase space $\rho-p_r$,  
(b) $\mathcal{S}_{\theta}$ for $\theta-p_{\theta}$,   
and (c) $\mathcal{S}_z$ for $\xi-p_z$,  
normalized by the corresponding initial values, 
obtained from the numerical calculation in Fig. 2.   
The red dashed lines represent exponential decay  
(a) $\exp(-t/T_r)$, 
(b) $f_{\theta} \exp(-t/T_{\theta})$
(c) $f_z \exp(-t/T_z)$
with fitted parameters 
$T_r=4.2$ ns, $T_{\theta}=T_z=2.2$ ns, $f_{\theta}=100$ and 
$f_z=10$. 
}
\label{fig4}
\end{figure}

\bibliography{refs.bib}

\end{document}